# Deformation mechanism for stabilization of long-range order in ferromagnetic polycrystals

A.A. Fraerman

*Institute for Physics of Microstructures, Russian Academy of Sciences,*

*Nizhny Novgorod 607680, Russian Federation*

*email: andr@ipmras.ru*

The influence of magnetostriction on static fluctuations of the magnetic moment in ferromagnetic polycrystals has been theoretically studied. Conditions have been found under which magnetoelastic interaction leads to stabilization of long-range magnetic order in these systems.

Ferromagnetic polycrystals are a broad class of materials with a variety of applications. Their magnetic properties depend on the distribution of crystallites in size and orientation and can differ significantly from the properties of single crystals of the same chemical composition. The question itself of the existence of long-range magnetic order in polycrystals is nontrivial. It is argued that in a polycrystal that does not have a preferred orientation in the distribution of crystallographic axes of crystallites, there is no long-range magnetic order. Magnetization is destroyed by static fluctuations in the orientation of the magnetic anisotropy axes, which are directly related to the orientation of the crystallographic axes in crystallites. The following arguments support this [1]. Let the direction of magnetization in a ferromagnetic polycrystal change on a scale $\sim L$. Then the exchange energy density due to this inhomogeneity has the value $\varepsilon_{ex} \sim \frac{J}{L^2}$, $J$ is the exchange constant. The magnitude of the anisotropy energy with a random distribution of crystallite axes $\varepsilon_a \sim -\frac{KM_0^2}{\sqrt{N}}$, $KM_0^2$ is the anisotropy energy density, $N$ is the number of crystallites in the region with linear dimensions $L$, $N \sim (L/l_c)^d$, $l_c$ is crystallite size, d is dimension of space. The minimum energy $\varepsilon = \varepsilon_{ex} + \varepsilon_a$ corresponds to the scale $L_c \sim l_w (l_w/l_c)^{d/(4-d)}$, where $l_w = \sqrt{J/KM_0^2}$ is the thickness of the domain wall. From this it follows that there is no long-range magnetic order in a polycrystal, and instead there is an inhomogeneous distribution of magnetization, called "magnetization ripples" [2]. The saturation field of a polycrystal $H_s$ is estimated as the field required for "exit" from the "magnetization ripple" state and strongly depends on the ratio of the crystallite size and the domain wall thickness $l_w$, $H_s \sim \frac{\varepsilon(L \to \infty) - \varepsilon(L = L_c)}{M_0} \sim KM_0 (l_c/l_w)^{2d/(4-d)}$, $M_0$ is saturation magnetization.

In the extensive literature devoted to the study of the magnetic properties of ferromagnetic polycrystals (see, for example, review [3]), the question of the possibility of stabilizing long-range order in these systems due to long-range magnetic interactions, such as magnetostatic and magnetoelastic interactions, is not discussed. The effect of stabilization of long-range order in two-dimensional ferromagnets with a degenerate ground state due to magnetostatic interaction is well known [4]. The question of the emergence of long-range order in two-dimensional magnets due to magnetoelastic coupling was also studied [5]. In this work, we show that magnetoelastic interactions can stabilize long-range order in polycrystals. The effect considered here is a static analogue of the "magnetoelastic gap" phenomenon, which manifests itself in the dynamics of magnets [6]. The article is organized as follows. In the first part, the main approximations are formulated and the terms that determine the energy of the system are written out. In the second part, the problem of free deformation of a three-dimensional polycrystal is considered and conditions are found under which magnetization fluctuations are small. In the third part, the problem of magnetization fluctuations in a thin polycrystal film elastically coupled to a non-magnetic substrate is discussed. The Conclusion provid esestimates of the effect considered.

## I. Problem formulation

In a ferromagnetic polycrystal, the components of the magnetocrystalline anisotropy tensor are random variables depending on the coordinates [2]. For uniaxial crystallites, the anisotropy energy density can be represented as

$$\varepsilon_a = -\frac{1}{2}K(n_i(\boldsymbol{r})M_i)^2, \qquad (1)$$

where $K$ is the anisotropy constant, $n_i(\boldsymbol{r})$ are the components of the random unit vector that determines the orientation of the crystallographic axes in the polycrystal. The statistical properties of this vector are considered known, $M_i$ are the components of the magnetization vector, which, generally speaking, are also random and depend on the coordinates. An additional exchange contribution to the energy is associated with the inhomogeneous distribution of magnetization, which we will present in the form

$$\varepsilon_{ex} = -l^2 M_i \Delta M_i \qquad (2)$$

$\Delta$ - Laplace operator, $l^2 = J/M_0^2$. The magnetization distribution $M_i$ induces stray fields with which magnetostatic energy is associated

$$\varepsilon_m = -\frac{1}{2}H_i M_i, \qquad (3)$$

where the stray field components $H_i$ satisfy the magnetostatic equations $\text{div}\mathbf{H} = -4\pi \text{div}\mathbf{M}$, $\text{rot}\mathbf{H} = 0$. The dependence of the energy of a polycrystal on deformations is determined by the sum of elastic and magnetoelastic energy

$$\varepsilon_e + \varepsilon_{me} = \frac{1}{2}\lambda U_{ll}^2 + \mu U_{ik}^2 + \beta U_{ik} M_i M_k, \tag{4}$$

$U_{ik} = \frac{1}{2}\left(\frac{\partial u_i}{\partial x_k} + \frac{\partial u_k}{\partial x_i}\right)$ is strain tensor, $u_i$ are components of the displacement vector. Lamé coefficients $\lambda$ and $\mu$ are related to Young's modulus ($E$) and Poisson's ratio ($\sigma$) by the relations $\lambda = \frac{E\sigma}{(1-2\sigma)(1+\sigma)}$, $\mu = \frac{E}{2(1+\sigma)}$ [7]. The dimensionless coefficient $\beta$ determines the magnitude of the magnetoelastic coupling. When recording energy (4), we neglect spatial fluctuations of elastic and magnetoelastic modules in a polycrystal and replace them with values averaged over crystallite orientations. This allows us to significantly simplify the problem and reduce it to the problem of magnetoelastic deformations in an elastically isotropic material. After differentiating (4) with respect to $U_{ik}$, we obtain the expression for the stress tensor

$$G_{ik} = \lambda U_{ll}\, \delta_{ik} + 2\mu U_{ik} + \beta M_i M_k, \tag{5}$$

$\delta_{ik}$ is the unit tensor. The displacement vector is determined from the equilibrium equations $\frac{\partial G_{ik}}{\partial x_k} = 0$. Let us imagine magnetization as the sum of its average value $M_i^{(0)}$ and fluctuations $m_i$

$$M_i = M_i^{(0)} + m_i(\mathbf{r}),\ M_i^{(0)} = M_0(1,0,0)$$

Let us assume that the fluctuations are small. Then from the condition of conservation of the magnetization modulus we have

$$m_i = \left(-\frac{1}{2M_0}(m_z^2 + m_y^2), m_y, m_z\right) \tag{6}$$

By limiting ourselves in expressions for energy (1)-(4) to terms that are quadratic in fluctuations, we can determine the mean square of fluctuations caused by the random distribution of anisotropy axes. The fulfillment of the condition of small transverse fluctuations compared to the average magnetization $\langle m_y^2 + m_z^2 \rangle < M_0^2$ is a necessary condition for the existence of long-range magnetic order in a polycrystal (brackets $\langle\ \rangle$ denote averaging over the orientations and sizes of crystallites). Next, we will consider the cases of a three-dimensional polycrystal and a thin polycrystalline film elastically coupled to the substrate.

## II. Three-dimensional polycrystal

With uniform magnetization of a three-dimensional polycrystal, uniform deformations occur (Fig. 1a). From the condition of no stress $G_{ik} = 0$, we find

$$U_{xx}^{(0)} = U_{yy}^{(0)} - \frac{\beta}{2\mu}M_0^2, U_{zz}^{(0)} = U_{yy}^{(0)} = \frac{\beta}{2\mu}M_0^2 \frac{\lambda}{3\lambda+2\mu} \tag{7}$$

Note that deformations (7) reduce the energy of the system, which becomes equal to

$$\varepsilon_0 = -\frac{\beta^2(\lambda+\mu)}{2\mu(3\lambda+2\mu)}M_0^4 = -2\frac{\beta^2}{E}M_0^4. \tag{8}$$

Let us represent the deformations as the sum of homogeneous deformations (7) and fluctuations caused by magnetization fluctuations $U_{ik} = U_{ik}^{(0)} + w_{ik} + v_{ik}$, where small deformations $w_{ik}$ are calculated linearly, and $v_{ik}$ in an approximation quadratic in magnetization fluctuations.

$$\varepsilon_e + \varepsilon_{me} = \varepsilon_0 + \beta w_{ik}\left(M_i^{(0)}m_k + M_k^{(0)}m_i\right) + \frac{1}{2}\lambda w_{ll}^2 + \mu w_{ik}^2 + \beta U_{ik}^{(0)}m_i m_k + G_{ik}^{(0)}w_{ik} + G_{ik}^{(0)}v_{ik}, \tag{9}$$

where $G_{ik}^{(0)}$ is the stress tensor in the zero fluctuation approximation. Since the extension is assumed to be free $G_{ik}^{(0)} = 0$, the last two terms in (9) are equal to zero. The term $\beta U_{ik}^{(0)}m_i m_k$ is the spontaneous anisotropy for magnetic fluctuations. Taking into account (7), we have

$$\beta U_{ik}^{(0)}m_i m_k = \frac{\beta^2(1+\sigma)}{E}M_0^2(m_y^2 + m_z^2) \tag{10}$$

Thus, the development of transverse magnetization fluctuations is energetically unfavorable and increases the energy of the system [6]. To answer the question of stabilization of long-range order, it is necessary to calculate the dispersion of magnetization fluctuations. The displacement vector $u_i$ satisfies the equilibrium equations

$$\Delta u_i + \frac{1}{1+2\sigma}\frac{\partial^2}{\partial x_i \partial x_k}u_k = -\frac{2\beta(1+\sigma)}{E}\frac{\partial}{\partial x_k}(M_i M_k) \tag{11}$$

The solution to this vector equation in an approximation linear in magnetization fluctuations is easy to find and for the Fourier image of the strain tensor we obtain

$$w_{ij} = -\frac{\beta(1+\sigma)}{E}\left[(\boldsymbol{m}(\boldsymbol{k})\boldsymbol{k})\left(M_i^{(0)}k_j + M_j^{(0)}k_i\right) - \frac{2}{1-\sigma}k_i k_j(\boldsymbol{m}(\boldsymbol{k})\boldsymbol{k})(\boldsymbol{M}^{(0)}\boldsymbol{k}) + (\boldsymbol{M}^{(0)}\boldsymbol{k})(m_i k_j + m_j k_i)\right] \tag{12}$$

where $k_j = \frac{q_j}{q}$, $q$ is the modulus of the wave vector. Substituting (12) into (9), we find the energy of the polycrystal associated with deformations. Assuming that the anisotropy constant $K$ is small, when calculating the anisotropy energy we can limit ourselves to the term linear in fluctuations

$$\varepsilon_a = -Kn_x M_0 (n_y m_y + n_z m_z) \tag{13}$$

Let us denote, $\eta_{xi}(r) = n_x(r)n_i(r), \eta_{xi}(q) = \int \eta_{xi}(r)\exp(-iqr)\, dr, i = z, y$. Since the direction of the vector $\boldsymbol{n}$ is uniformly distributed in the solid angle $4\pi$, $\langle \eta_{xi}(\boldsymbol{q})\eta_{xj}(-\boldsymbol{q})\rangle = \eta(q)\delta_{ij}$. Assuming the crystallite size distribution to be statistically homogeneous and isotropic, we write

$$\eta(q) = \int \langle n_x(r)n_i(r)n_x(r_1)n_i(r_1)\rangle \exp(-i\boldsymbol{q}(r-r_1))\, dr dr_1 = \frac{V}{15}\int \psi(\rho)\exp(i\boldsymbol{q}\boldsymbol{\rho})\, d\boldsymbol{\rho}, \tag{14}$$

where $V$ is the volume of the polycrystal. The correlation function $\psi(\rho)$ defines the region of space in which the vector $\boldsymbol{n}$ does not change its direction. For a statistically homogeneous and isotropic distribution, this function can be chosen in the form $\psi(\rho) = \exp(-k_c^2 \rho^2)$, then the crystallite size is $l_c \sim 1/k_c$. The Fourier image of the correlator (14) depends only on the modulus of the wave vector and is equal to

$$\eta(q) = V \frac{\pi\sqrt{\pi}}{15} k_c^{-3} \exp(-q^2/4k_c^2) \tag{15}$$

Let us represent magnetostatic energy (3) in the form

$$\varepsilon_m = 2\pi \int \frac{1}{q^2} |\boldsymbol{q}\boldsymbol{m}|^2 \, d\boldsymbol{q} \tag{16}$$

Substituting the solutions found, for the energy of the polycrystal we finally obtain

$$\varepsilon = \int \left( f_1 |\boldsymbol{m}|^2 + f_2 |\boldsymbol{m}\boldsymbol{k}|^2 - KM_0 \eta_{xi}(\boldsymbol{q}) m_i(-\boldsymbol{q}) \right) d\boldsymbol{q}, \boldsymbol{k} = \frac{\boldsymbol{q}}{q}, i = z, y \tag{17}$$

where

$$f_1 = l^2 q^2 + \frac{\beta^2(1+\sigma)}{E} M_0^2 (1 - k_x^2) \tag{18}$$

$$f_2 = \frac{\beta^2(1+\sigma)}{E} M_0^2 \left( \frac{2}{1-\sigma} k_x^2 - 1 \right) + 2\pi \tag{19}$$

Equating the energy variation (17) with respect to $m_i(-\boldsymbol{q})$ to zero, we obtain a system of equations for determining fluctuations

$$f_1 m_i + f_2 (\boldsymbol{mk}) k_i = K M_0 \eta_{xi}(\boldsymbol{q}) \tag{20}$$

Solving this system, we find the average square of transverse fluctuations over the distribution of random anisotropy

$$\langle m_\perp^2 \rangle = \langle m_x^2 + m_y^2 \rangle = \frac{1}{V} 2\pi K^2 M_0^2 \iint \left[ \frac{1}{f_1^2} + \frac{1}{(f_1 + f_2(1-k_z^2))^2} \right] \eta(q) q^2 \sin\theta \, d\theta \, d, \tag{21}$$

$\theta$ - the angle between the wave vector and the oX axis. Details of the calculation of the integrals in (21) are given in the Appendix. The magnitude of fluctuations is determined by the ratio of the crystallite size $l_c$ and the characteristic length $l_m$ associated with the spontaneous strain anisotropy constant $b = \frac{\beta^2(1+\sigma)}{E} M_0^2$, $l_m^2 = \frac{l^2}{b}$. Let us present expressions for the squared fluctuations of the magnetic moment in limiting cases. If the crystallite size is large $k_c l_m \ll 1$, then

$$\langle m_\perp^2 \rangle = \frac{2\pi}{15} \frac{K}{b} \frac{1}{k_c^2 l_w^2} M_0^2, \; l_w^2 = l^2/K \tag{22}$$

In the opposite limiting case $k_c l_m \gg 1$, we have

$$\langle m_\perp^2 \rangle = \frac{\pi^4 \sqrt{\pi}}{30} \frac{K}{b} \frac{1}{k_c^3 l_w^2 l_m} M_0^2, \tag{23}$$

The condition for small fluctuations and stabilization of long-range order is the condition $\langle m_\perp^2 \rangle \ll M_0^2$.

### III. Ferromagnetic film on a substrate

Let us consider a ferromagnetic film on a non-magnetic substrate, which would be perfectly matched without taking into account the magnetization of the film. We will assume that magnetization spontaneously arose in the film, the direction of which will be chosen as the *oX* axis. This magnetization, due to magnetoelastic coupling, leads to film deformation, which is prevented by the substrate. Assuming that the film and substrate are elastically connected, non-uniform deformation leads to bending of the entire system, and the shape of the plate is given by the equation $z = \xi(x, y)$ (Fig. 1b). Let us find the deformations that arise in this case. Assuming the substrate is sufficiently thin, we will assume that the normal components of the strain tensor (5) are equal to zero $G_{iz} = 0$ [7]. It follows from this that the components of the strain tensor are equal to zero

$$U_{xz} = U_{yz} = 0; \; \frac{\partial u_x}{\partial z} = -\frac{\partial u_z}{\partial x}, \frac{\partial u_y}{\partial z} = -\frac{\partial u_z}{\partial y} \tag{24}$$

Assuming that $u_z = \xi$ and integrating equations (24) over Z, we find $u_x = -(z - z_0)\frac{\partial \xi}{\partial x}, u_y = -(z - z_0)\frac{\partial \xi}{\partial y}$. Here $z_0$ is the coordinate of the neutral line separating the areas of tension and compression of the plate. We will find both quantities, $\xi$ and $z_0$, from the condition of minimum energy. Since the source of deformation - magnetization - acts only along the X axis, we will assume that $\xi = \xi(x)$. In this case, $U_{xx}$ and $U_{zz} = -\frac{\lambda}{\lambda+2\mu}U_{xx}$ are non-zero components of the strain tensor. After integration along the Z axis, taking into account the fact that magnetization exists only in the film, and the elastic modules of the film and substrate are equal, we find

$$\varepsilon_e + \varepsilon_{me} = \frac{E}{2(1-\sigma^2)}\left(\frac{H^3}{12} + z_0^2 H\right)\left(\frac{\partial^2 \xi}{\partial x^2}\right)^2 - \beta M_0^2 \left(\frac{1}{2}Hh - \frac{1}{2}h^2 - z_0 h\right)\left(\frac{\partial^2 \xi}{\partial x^2}\right), \tag{25}$$

where $h$ is the film thickness, $H$ is the total thickness of the film and substrate. The minimum energy (25) corresponds to the values

$$z_0 = -\frac{1}{6}\frac{H^2}{H-h}, \frac{\partial^2 \xi}{\partial x^2} = -\frac{(1-\sigma^2)\beta h}{Ez_0 H}M_0^2 = 6\frac{(1-\sigma^2)\beta h(H-h)}{EH^3}M_0^2 \tag{26}$$

Thus, the system takes the shape of a parabolic cylinder, and the neutral line is shifted from the center by an amount $z_0$. If the thickness of the substrate becomes small ($h \to H$), the curvature $\frac{\partial^2 \xi}{\partial x^2}$ tends to zero, and the average deformation over the film thickness approaches a constant value $U_{xx} \to -\frac{(1-\sigma^2)\beta M_0^2}{E}$. In the opposite limiting case of a thin film ($h \ll H$), the gain in energy density due to deformations is equal to $\varepsilon_0 = -2\frac{\beta^2 M_0^4 (1-\sigma^2)h}{EH}$, which is $H/h$ times less gain in the case of free deformation (8).

## Conclusion

So, the condition for the smallness of fluctuations caused by the spread of the anisotropy axes compared to the average spontaneous magnetization is the inequality $\langle m_\perp^2 \rangle \ll M_0^2$. This condition can only be satisfied if magnetostriction is taken into account. If the magnetostriction coefficient $\beta \to 0$, then the value of the "magnetoelastic gap" $b \to 0$ tends to zero, and therefore the amplitude of fluctuations becomes infinitely large. Our assumption about the existence of average spontaneous magnetization is not correct, which indirectly confirms the existing statement about the absence of long-range magnetic order in a ferromagnetic polycrystal. When taking into account the magnetoelastic coupling, the situation changes qualitatively. The intensity of fluctuations depends on the ratio of the size of the crystalline block $l_c \left(k_c = \frac{2\pi}{l_c}\right)$ and the characteristic "magnetostrictive" length $l_m = l/\sqrt{b}$. The size of the crystalline block in films of ferromagnetic metals is $l_c \sim 10 - 100$ nm (see, for example, [9]), the length $l \sim \sqrt{J/M_0^2}$ is taken equal to 10 nm. The value of the "magnetoelastic gap" for the free deformation of a ferromagnet is equal to $b = \frac{\beta^2(1+\sigma)}{E}M_0^2$. If we assume that the elastic modulus $\sim 10^{11} - 10^{12}$ erg/cm³, $M_0^2 \sim 10^6$ erg/cm³, $\beta \sim 1$, then $b \sim 10^{-5} - 10^{-6}$ [8]. Then the "magnetostriction length" $l_m \sim 300 - 1000$ nm, as a rule, is larger than the crystallite size. In this case, to estimate the intensity of fluctuations, one should use formula (23), which can be rewritten as the condition $\frac{\pi}{240}\frac{\sqrt{\pi}}{\sqrt{b}}\frac{K^2}{1}\frac{l_c^3}{l^3} < 1$. Taking into account that K~0.1-1, this condition can be satisfied for crystallites of small sizes $l_c < l$.

Thus, stabilization of long-range order in ferromagnetic polycrystals is possible. This stabilization is due to a decrease in the energy of the magnetized ferromagnet during deformation, which is accompanied by the formation of a "magnetoelastic gap" for magnetization fluctuations caused by fluctuations in the orientation of the crystalline axes in the polycrystal. In a thin film elastically connected to the substrate, magnetoelastic deformations are virtually absent. Indeed, the average deformation over the film thickness $U_{xx} \approx -\frac{(1-\sigma^2)\beta M_0^2 h}{EH}$ does not exceed a small value $\sim 10^{-10}$ ($h \sim 100$ nm, $H \sim 0.1$ cm). Those, the film is in a stressed state and a "magnetoelastic gap" does not appear. However, relaxation of the stressed state can occur in this system [9], just as stress relaxation occurs in epitaxial films due to the formation of misfit dislocations [10]. Analysis of these processes and their influence on long-range magnetic order in ferromagnetic polycrystals is beyond the scope of this work.

I thank M.V. Sapozhnikov and D.A. Tatarsky for useful comments, M.A. Kuznetsov and R.V. Gorev for their help in preparing the manuscript for publication.

**Appendix**

Let's calculate the integrals in formula (21)

$$\langle m_\perp^2 = m_x^2 + m_y^2 \rangle = \frac{1}{V} 2\pi K^2 M_0^2 (I_1 + I_2)$$

$$I_1 = \iint \left[\frac{1}{f_1^2}\right] \eta(q) q^2 Sin\theta d\theta dq$$

$$I_2 = \iint \left[\frac{1}{(f_1 + f_2(1-k_z^2))^2}\right] \eta(q) q^2 Sin\theta d\theta dq$$

In the integral over the angular variable, we make the replacement $= Cos\theta$. Then the expression for $I_1$ will take the form

$$I_1 = \frac{\pi\sqrt{\pi}}{15} k_c^{-3} \int_0^\infty q^2 \exp(-q^2/4k_c^2) \int_{-1}^{1} \frac{dx}{[(lq)^2 + b - bx^2]^2} \tag{A1}$$

Introducing the notation for deformation anisotropy $b = \frac{\beta^2(1+\sigma)}{E} M_0^2$ and the associated characteristic length $l_m^2 = l^2/b$, we obtain

$$\int_{-1}^{1} \frac{dx}{[(lq)^2 + b - bx^2]^2} = \frac{((l_m q)^2 + 1)^{1/2}}{b^2((l_m q)^2 + 1)^2} \left[\frac{((l_m q)^2 + 1)}{(l_m q)^2((l_m q)^2 + 1)^{1/2}} + \frac{1}{2} ln\left(\frac{((l_m q)^2 + 1)^{1/2} + 1}{((l_m q)^2 + 1)^{1/2} - 1}\right)\right] \tag{A2}$$

Then

$$I_1 = \frac{\pi\sqrt{\pi}}{15b^2} k_c^{-3} \int_0^\infty q^2 \exp\left(-\frac{q^2}{4k_c^2}\right) \left[\frac{1}{((l_m q)^2 + 1)(l_m q)^2} + \frac{1}{2((l_m q)^2 + 1)^{\frac{3}{2}}} ln\left(\frac{((l_m q)^2 + 1)^{\frac{1}{2}} + 1}{((l_m q)^2 + 1)^{\frac{1}{2}} - 1}\right)\right] dq$$

$$= \frac{\pi\sqrt{\pi}}{15b^2 l_m^2} k_c^{-3} \int_0^\infty \exp\left(-\frac{q^2}{4k_c^2}\right) \left[\frac{1}{((l_m q)^2 + 1)} + \frac{(l_m q)^2}{2((l_m q)^2 + 1)^{\frac{3}{2}}} ln\left(\frac{((l_m q)^2 + 1)^{\frac{1}{2}} + 1}{((l_m q)^2 + 1)^{\frac{1}{2}} - 1}\right)\right] dq \tag{A3}$$

Let's consider limiting cases.

1. Crystallite size is large $k_c l_m \ll 1$

$$I_1 = \frac{\pi\sqrt{\pi}}{15b^2 l_m^2} k_c^{-3} \int_0^\infty \exp\left(-\frac{q^2}{4k_c^2}\right)\left[1 + \frac{1}{2}(l_m q)^2 \ln\left(\frac{4}{(l_m q)^2}\right)\right] dq \approx \frac{\pi^2}{15 b^2 l_m^2} k_c^{-2}$$

(A4)

Neglecting the second term in parentheses, for the mean square of magnetization fluctuations, we have

$$\langle m_\perp^2 \rangle = \frac{2\pi^3}{15} \frac{K}{b} \frac{1}{k_c^2 l_{dw}^2} M_0^2, \quad l_{dw}^2 = l^2/K \quad (A5)$$

2. The crystallite size is small $k_c l_m \gg 1$

$$I_1 = \frac{\pi\sqrt{\pi}}{15 b^2 l_m^2} k_c^{-3} \int_0^\infty \left[\frac{1}{((l_m q)^2+1)} + \frac{(l_m q)^2}{2((l_m q)^2+1)^{\frac{3}{2}}} \ln\left(\frac{((l_m q)^2+1)^{\frac{1}{2}}+1}{((l_m q)^2+1)^{\frac{1}{2}}-1}\right)\right] dq = \frac{\pi\sqrt{\pi}}{15 b^2 l_m^3} k_c^{-3} C$$

$$C = \int_0^\infty \left[\frac{1}{(x^2+1)} + \frac{x^2}{2(x^2+1)^{\frac{3}{2}}} \ln\left(\frac{(x^2+1)^{\frac{1}{2}}+1}{(x^2+1)^{\frac{1}{2}}-1}\right)\right] dx = \frac{\pi^2}{4}$$

In this approximation, the contribution to transverse fluctuations is equal to

$$\langle m_\perp^2 \rangle = \frac{\pi^4 \sqrt{\pi}}{30} \frac{K}{b} \frac{1}{k_c^3 l_w^2 l_m} M_0^2, \qquad (A6)$$

$$l_w^2 = J/K$$

Let us show that the corrections to expressions (A5) and (A6) due to $I_2$ are small

$$f_1 + f_2(1 - k_z^2) = l^2 q^2 + b(1 - k_z^2) + b\left(\frac{2}{1-\sigma} k_z^2 - 1\right)(1 - k_z^2) + 2\pi(1 - k_z^2)$$

$$= l^2 q^2 + b\frac{2}{1-\sigma} k_z^2 (1 - k_z^2) + 2\pi(1 - k_z^2)$$

Since $b\frac{2}{1-\sigma} k_z^2 \ll 2\pi$, the denominator in $I_2$ takes on a form similar to the denominator in $I_1$ with the replacement $\leftrightarrow 2\pi$. Thus, the contribution from $I_2$ is $1/b$ times less than from $I_1$, and we will neglect it.

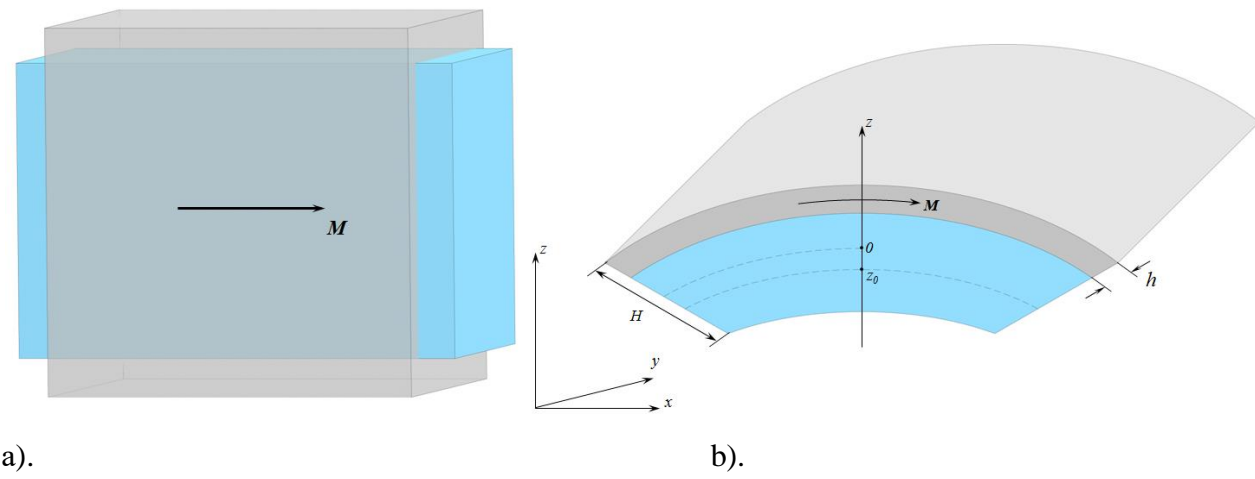

Fig.1 a). free deformation of a ferromagnet during magnetization, b). deformation during magnetization of a thin film elastically connected to the substrate